\documentstyle[12pt]{article}

\font\eightrm=cmr8
\font\tenrm=cmr10

\newcommand{\bref}[1]{\,(\ref{#1})} 
\newcommand{\ct}[1]{\cite{#1}} 

\newcommand{\be}{\begin{equation}} 
\newcommand{\ee}{\end{equation}}

\newcommand{\beq}{\begin{equation}}
\newcommand{\eeq}{\end{equation}}

\def\theequation{\thesection.\arabic{equation}}
\def\@eqnnum{{\rm (\theequation)}}

\def\lsim{\mathrel{\rlap{\lower4pt\hbox{\hskip1pt$\sim$}}
    \raise1pt\hbox{$<$}}}
\def\gsim{\mathrel{\rlap{\lower4pt\hbox{\hskip1pt$\sim$}}
    \raise1pt\hbox{$>$}}}
\def\frac#1#2{{{#1} \over{#2}}}

\begin{document}
\begin{titlepage}

\begin{flushright}
{LBNL-55463\\ }
{\hfill August 2004 \\ }
\end{flushright}
\vglue 0.2cm

\begin{center}
{
{\tenrm{
{\bf
{ON THE SPEED OF GRAVITY AND \\ THE JUPITER/QUASAR MEASUREMENT \\ }} 
}}
\vglue 1.0cm
{\eightrm{
{STUART SAMUEL$^{1}$ \\ } 
}}
\vglue 0.5cm 

\vglue 0.4cm
{\eightrm{
{\it Theory Group, MS 50A-5101 \\}
{\it Lawrence Berkeley National Laboratory\\}
{\it One Cyclotron Road\\}
{\it Berkeley, CA 94720 USA\\}
$^{1}${\it{samuel@thsrv.lbl.gov}} \\ 
}} 

\vglue 0.8cm


{\bf Abstract}
}
\end{center} 
{\rightskip=3pc\leftskip=3pc
\quad I present the theory and analysis behind 
the experiment by Fomalont and Kopeikin 
involving Jupiter and quasar J0842+1845 
that purported to measure the speed of gravity. 
The computation of the $v_J/c$ correction 
to the gravitational time 
delay difference relevant to the experiment is derived, 
where $v_J$ is the speed of Jupiter as measured from Earth.
Since the $v_J/c$ corrections are too small to have been measured 
in the Jupiter/quasar experiment, 
it is impossible that the speed of gravity was extracted from the data, 
and I explain what when wrong with the data analysis. 
Finally, 
mistakes are shown in papers by Fomalont and Kopeikin 
intended to rebut my work and the work of others. 


\medskip 
\medskip 
\noindent
{\it {PACS indices:}} 04.20.Cv, 04.80.Cc, 04.25.Nx, 98.54.Aj, 96.30.Kf
}

\vfill        
\eject
\end{titlepage}

\newpage

\baselineskip=20pt

{\bf\large\noindent 1.\ Introduction}\vglue 0.2cm
\setcounter{section}{1}   
\setcounter{equation}{0}  

Albert Einstein constructed his general theory of relativity 
so that classical gravity would be compatible with the principles of special relativity. 
As such, 
gravitational waves and the influences of gravity 
are suppose to propagate at the speed of light $c$. 
For example, 
if hypothetically the Sun were to explode into two pieces 
then the force of gravity on the Earth would change. 
However, it would not happen suddenly but 8 and 1/3 minutes later,  
since this is the time it takes gravitational effects (and light) 
to travel from the Sun to the Earth. 
In other words,
the Earth would continue in its almost circular orbit 
as governed by the gravity of a single massive central body 
for another 8 and 1/3 minutes, 
only after which would its motion be determined by the two exploding pieces. 
Since gravitational waves have not yet been detected, 
it has not been possible to test 
whether they travel at the speed $c$, 
nor has there been a system in which the speed of propagation 
of gravitational influences $c_g$ has been measured. 

The lack of a measurement of $c_g$ inspired S.\ Kopeikin 
to propose an experiment to test whether $c_g$ is the speed of light.\ct{kopeikin0105060} 
The ideas behind his proposal are outlined in the next four paragraphs. 

When electromagnetic waves pass by a massive object $M$, 
two effects occur: 
Firstly, the waves are very slighly bent, 
and, secondly, there is a tiny delay in the transmission time. 
Both are prominent effects of general relativity, 
and the latter is known 
as the Shapiro time delay.\ct{shapiro,shapiroetal68,shapiroetal71} 

The physical solution of Einstein's equations involve 
the position $\vec x_M $ of gravity-generating objects at retarded times: 
\be
t_{ret} = t - \vert \vec x - \vec x_M (t_{ret}) \vert/c
\,.
\label{retardtimegen}
\ee
Here, $\vec x$ is the location at which gravity is exerting its influence. 
The use of retarded times $t_{ret}$ as given 
in Eq.\bref{retardtimegen}  
implies that the effect of gravity propagates at $c$. 
To allow for gravity to propagate at a different speed $c_g$, 
one would expect to replace Eq.\bref{retardtimegen} by 
\be 
t_{ret} = t - \vert \vec x - \vec x_M (t_{ret}) \vert/c_g 
\,.
\label{retardtimemod}
\ee

On September 8, 2002, 
a conjunction of Jupiter and quasar J0842+1845 took place. 
Kopeikin argued that this event could be used 
to measure the speed of propagation of gravity.\ct{kopeikin0105060} 
The thinking behind his proposal originates from the previous paragraph. 
Being very far away, the location of quasar J0842+1845 in the sky is virtually fixed. 
Jupiter, however, moves. 
Its position at the retarded time depends sensitively 
on its velocity and on $c_g$ through Eq.\bref{retardtimemod}. 
Therefore, one {\it {might}} expect that a precise measurement of 
the Shapiro time delay due to Jupiter on the signal from quasar J0842+1845 
would permit a determination of $c_g$. 

The possible effect of Jupiter's velocity $\vec v_J$ on the Shapiro time delay 
can be illustrated from an example: 
Suppose that Jupiter is moving toward the direction of the quasar waves. 
If $c_g$ were infinite, which corresponds to the Newtonian limit of general relativity, 
then it would {\it {appear}} that the instantaneous position of Jupiter would be relevant. 
If $c_g = c$, then the position of Jupiter evaluated at the retarded time 
is somewhat farther away than the instantaneous position, 
and if $c_g < c$, then the position as determined 
by Eq.\bref{retardtimemod} is even further away. 
See Figure 1. 

Using an array of radio telescopes that stretched 
across the United States all the way to Germany, 
E.\,Fomalont and S.\,Kopeikin measured the tiny Shapiro time delays. 
By applying Very Long Baseline Interferometry (VLBI), 
they achieved a remarkable sensitivity at the picosecond level. 
The results for $c_g$, 
which were announced in January 2003
meeting of the American Astronomical Society 
held in Seattle, Washington, 
immediately caught the attention of the media. 
The New York Times, for example, 
featured an article entitled 
``Einstein Was Right on Gravity's Speed.''\ct{nyt}  
At the meeting, 
Fomalont and Kopeikin announced that $c_g = c$ to within 20\% 
and this result was subsequently published.\ct{fomalontkopeikin} 

If this result is correct 
then it is a fundamental confirmation 
of Einstein's general theory of relativity.
However, 
shortly after the American Astronomical Society meeting, 
a debate among astrophysicists arose about the Jupiter/quasar experiment. 
Several papers appeared arguing that Fomalont and Kopeikin 
had not accomplished their goal of measuring $c_g$. 
H.\,Asada had published an early work\ct{asada} 
stating that Kopeikin's idea actually measures 
the speed of light instead of the speed of gravity, 
C. M. Will also argued that the measurements were not directly sensitive to $c_g$,\ct{will} 
and other papers also appeared criticizing the theory behind the experiment.\ct{faber,asada0308343,carlip}
The difficulty that many of the above works were addressing 
is that there is no agreed upon method for extending 
Einstein's theory to the case for which $c_g \ne c$. 
Kopeikin has repeatly tried to defend these criticisms 
of his work.\ct{kopeikin0212121,kopeikin0310059} 

In a Physical Review Letter\ct{samuelprl}  
(hereafter referred to as my Physical Review Letter), 
I bypassed the issue of how to extend general relativity to the case 
where the speed of gravity does not equal the speed of light 
and computed the $v_J/c$ corrections to the Shapiro time delay 
in Einstein's theory 
using a relatively simple method. 
The $v_J/c$ corrections did not agree with those derived 
by Kopeikin\ct{kopeikin0105060,kopeikin0212121} 
when his parameter $c_g$ was set to $c$. 
The correct theoretical formula 
implied that the $v_J/c$ dependence 
was at least 100 times smaller 
than could have been measured by the array of radio telescopes 
used in the Fomalont/Kopeikin experiment. 
In other words, 
the speed of gravity could not have been extracted 
from the Jupiter/quasar measurement. 
My Physical Review Letter definitively 
settled the speed of gravity controversy: 
The parameter $c_g$ has not been measured. 

Kopeikin had argued that there is an enhancement 
in the $v_J/c$ correction to the Shapiro time delay by a factor 
of $1/\theta$ where $\theta$ is the angle between Jupiter 
and the quasar.\ct{kopeikin0105060,kopeikin0212121} 
Since $\theta$ is small,  
it would seem to be that the $v_J/c$ correction is sizeable. 
However, 
no such $1/\theta$ enhancement is actually present. 

My Physical Review Letter pinpointed the source of the discrepancy. 
The leading-order, velocity-independent part of the term that 
Fomalont and Kopeikin measured 
depends on the distance $\xi$ of closest approach 
of the radios waves to Jupiter as $1/\xi$. 
This distance is determined by the position of Jupiter 
and the radio waves as the latter pass by Jupiter. 
If $R_{EJ}$ and $\theta_{obs}$ respectively denote 
the Earth-Jupiter distance and the angle that an astronomer 
observes between Jupiter and the quasar, 
then 
\be 
\xi = R_{EJ} \theta_{obs} 
\,.
\label{xitheta}
\ee 
The angle $\theta_{obs}$ is the one determined by the geometry 
of the positions of Jupiter, the quasar and the Earth {\it {at the time 
in which the radio waves pass by Jupiter}}. 

Fomalont and Kopeikin parametrized their data in terms 
of an angle $\theta_1$ determined by the geometry 
of the positions of the above three objects {\it {at the time 
in which the radio waves arrived on Earth}}. 
These two angles differ: 
During the time in which the quasar signals travel from Jupiter to Earth, 
Jupiter moves a significant distance. 
The relation between the two angles is 
\be
   \theta_{obs} \approx \theta_{1} + { {\vec n \cdot \vec v_J } \over {c} }
\,,
\label{thetaobstheta1}
\ee 
where $\vec n$ is a unit vector pointing from Jupiter 
to the quasar's radio waves at the time of closest approach. 
When the factor $1/\xi$ appearing 
in the leading velocity-independent term 
is expressed in terms of $\theta_{1}$, 
it appears to become velocity dependent because 
\be 
  {{1} \over {\xi}} = {{1} \over {R_{EJ} \theta_{obs}}} \approx 
  {{1} \over {R_{EJ} \theta_1}} \left( 1 - 
   {{\vec n \cdot \vec v_J} \over {c \theta_1}} \right) 
\,.
\label{xiexp}
\ee 
In summary, 
parametrizing the Shapiro time delay using $\theta_1$ 
makes the leading term seem to depend on the velocity of Jupiter. 
Furthermore, 
the fictitious velocity-dependent term appears enhanced. 

Fomalont and Kopeikin took their data 
and fit them to the leading order term but 
parametrized them in terms of $\theta_1$. 
They then extracted the $v_J/c$ dependence 
calling it the $v_J/c_g$ correction. 
Given this procedure and that the data have error bars, 
it is not surprising that such a procedure produced 
the purported result that $c_g = c$ to within 20\%. 
The conclusion that the speed of gravity is the speed of light 
to within experimental errors is not valid due to faulty analysis 
and a flaw in the theoretical understanding of the situation. 

Since the difference between $\theta_{obs}$ and $\theta_{1}$ 
is due to the change in the position of Jupiter 
as the quasar's radio waves propagate from Jupiter to Earth, 
it is clear that the parameter $c$ 
in Eq.\bref{xiexp} 
is the speed of light and has nothing to do with the speed of gravity. 
Therefore there was no justification in using $c_g$ in lieu of $c$ 
when expressing the leading term in terms of $\theta_1$. 

The measurement of Fomalont and Kopeikin 
of the Shapiro time delay due to Jupiter 
is a remarkable experimental achievement. 
One should remember that 
the non-Newtonian effects of general relativity 
due to a planet had hitherto never been detected. 
However, the experiment has 
little theoretical significance or fundamental importance, 
and it indicates nothing about the speed of gravity. 

Most of our notation conforms to that of 
references \ct{kopeikin0105060}, \ct{kopeikin0212121} and \ct{samuelprl}. 
There are several small dimensionless parameters 
characterizing the Jupiter/quasar measurement: 
$ { {G_N M_J } / ( {\xi c^2} ) } \approx 6 \times 10^{-9}$, 
where $G_N$ and $M_J$ are respectively 
Newton's constant and the mass of Jupiter; 
${ {v_J}/ {c} } \sim 10^{-4}$;
${ {B} / {\xi} } \le 0.006$ 
where $B$ is the distance between any two VLBI stations on Earth; 
and $\theta_{obs} = { {\xi} / { R_{EJ} } } \sim 0.001 $. 
These parameters, in the order given above, 
respectively represent 
the weakness of Jupiter's gravity on the radio waves, 
the non-relativistic nature of Jupiter's motion, 
the diminutive size of Earth compared to Jupiter, 
and the small observational angle between Jupiter and the quasar. 

In our analysis, 
we neglect the square of any of the above quantities. 
In particular, 
relativistic and high-order gravitational effects are ignored.

\medskip
{\bf\large\noindent 2.\ The Leading Order Result}\vglue 0.2cm
\setcounter{section}{2}   
\setcounter{equation}{0}  

Suppose that Jupiter is not moving.  
We refer to this as the static situation. 
Then, the Shapiro time delay 
for an electromagnetic wave travelling from the quasar 
past Jupiter to Earth is 
\be
  \Delta t = 
  {{2G_N M_J} \over {c^3}}
  \left( {1+\ln \left( {{{4R_{JQ} R_{EJ}} \over {\xi^2}}} \right)} \right) 
\,,
\label{shapirotimedelay} 
\ee
where $R_{JQ}$ is the distance from Jupiter to the quasar. 
Eq.\bref{shapirotimedelay} is a textbook result.\ct{weinberg} 

In the quasar/Jupiter experiment, 
a series of radio telescopes detected the quasar signals 
during the conjunction.
The time difference $\Delta \left( t_1 , t_2 \right)$ 
between two such Shapiro delays $\Delta t_2$ and $\Delta t_1$ was measured: 
\be 
\Delta \left( t_1 , t_2 \right) = \Delta t_2 - \Delta t_1
\,.
\label{deltadef}
\ee 

The experimental situation is shown in Figure 2. 
Signals 1 and 2 propagate from the quasar past Jupiter and arrive 
at times $t_1$ and $t_2$ on Earth at detectors located at positions 
$ \vec x_1 (t_1)$  and $ \vec x_2 (t_2)$. 

Using Eq.\bref{shapirotimedelay} in Eq.\bref{deltadef} yields 
\be
  \Delta \left( {t_1,t_2} \right) = 
  \Delta t_2 - \Delta t_1 = 
  {{2 G_N M_J} \over {c^3}}\ln \left( {{{r_{2J} \xi_1^2} \over {r_{1J} \xi_2^2}}} \right) 
    \approx {{4G_N M_J \Delta \xi } \over {\xi c^3}} 
\,, 
\label{deltalo0}
\ee 
where $\Delta \xi = \xi_1 - \xi_2$ 
and $r_{1J}$ (respectively, $r_{2J}$) is the distance 
between the first (respectively, second) detector and Jupiter. 
The last equality 
in Eq.\bref{deltalo0} 
follows because 
$\Delta \xi$ is significantly smaller than either $\xi_1$ or $\xi_2$,  
and the differences in the distances 
$r_{1J}$ and $r_{2J}$ between Jupiter and detector can be neglected. 
We use $\xi$ without a subscript 
to denote either of the detector-specific impact parameters $\xi_1$ or $\xi_2$ 
when the distinction between the two is not important. 
Although gravitational effects are often long-ranged, 
the Shapiro time delay difference $\Delta \left( t_1 , t_2 \right)$ 
is generated in the vicinity of Jupiter 
as is evident from Eq.\bref{deltalo0}: 
$\Delta \left( t_1 , t_2 \right)$ depends only on impact parameters. 

It is convenient to express 
$\Delta \xi = \xi_1 - \xi_2$ in terms of the displacement  
between the two detectors $\vec B = \vec x_2 (t_2) - \vec x_1 (t_1)$ 
because $\vec B$ is easily determined experimentally.
See Figure 2. 
The radio signals from the quasar 
are bent slightly by an amount $\Delta \varphi$ as they pass by Jupiter.  
However, it turns out that this effect can be neglected  
as we now show. 

The bending of a single wave is 
given by\ct{weinberg} 
\be 
  \Delta \varphi = {{4G_N M_J} \over {\xi c^2}}
\,.
\label{grbending}
\ee  
The angle that eventually arises between the two rays is
\be  
  \delta \Delta \varphi = \Delta \varphi_2 - \Delta \varphi_1 
    = { {4G_N M_J \Delta \xi } \over {\xi^2 c^2} }
\,.
\label{tworayangle}
\ee 
Since the separation between the rays starts as $\Delta \xi$ 
and increases as the distance times $\delta \Delta \varphi$, 
\be 
  -\vec n \cdot \vec B = \Delta \xi + R_{EJ} \delta \Delta \varphi 
  = \Delta \xi \left( {1+{{4G_N M_J R_{EJ}} \over {\xi^2 c^2}}} \right) 
  \approx \Delta \xi  
\,.
\label{bxi} 
\ee  
The last equality follows because 
$$ 
  {{4G_NM_JR_{EJ}} \over {\xi^2 c^2}} 
   \le {{4G_NM_JR_{EJ}} \over {R_J^2c^2}} \sim 0.001 
\,,
$$  
where $R_J$ is the radius of Jupiter.  
Since the angular deflection caused by Jupiter is so small,  
the separation between the two rays remains essentially constant. 
Indeed, 
including the second term 
in Eq.\bref{bxi} in our analysis below 
only leads to corrections proportional to Newton's constant squared. 

Substituting Eq.\bref{bxi} into \bref{deltalo0},  
one obtains the following for the static situation  
\be
  \Delta \left( {t_1, t_2} \right) 
  = -{{4G_N M_J\vec n\cdot \vec B} \over {\theta_{obs} R_{EJ} c^3}} 
\,,
\label{deltalo}
\ee 
where we have used $\xi = \theta_{obs} R_{EJ}$.   
Eq.\bref{deltalo} is expressed in quantities measurable on Earth. 

\medskip
{\bf\large\noindent 3.\ The $v_J/c$ Corrections}\vglue 0.2cm
\setcounter{section}{3}   
\setcounter{equation}{0}  

Now consider the case in which Jupiter is moving with a velocity 
$\vec v_J$ with respect to the Earth. 
In principle, 
one should start with Einstein's equation 
\be 
  R^{\mu \nu} - {{1} \over {4}} g^{\mu \nu} R 
    = - { {8 \pi G_N} \over {c^2} } T^{\mu \nu} 
\,, 
\ee 
in which $R_{\mu \nu}$, the curvature tensor, 
is related to the stress-energy tensor $T^{\mu \nu}$. 

For a static Jupiter, 
the stress-energy tensor only has a ``00'' component: 
$T^{00} = \rho_J (x)$, where $\rho_J$ is Jupiter's mass density. 
The metric $g_{\mu \nu}$ 
in spherical-like coordinates about the center of Jupiter 
is given by the Schwarzschild solution 
\be 
 c^2 d\tau^2 = \left( { 1 - {{2 M_J G_N} \over {r c^2}} } \right) c^2 dt^2 
  - \left( { 1- {{2 M_J G_N} \over {r c^2}} } \right)^{-1} dr^2 
  - r^2 \theta^2 -r^2 \sin^2 \theta d \varphi^2 
\,, 
\label{schwarzschild} 
\ee 
a result that is only valid exterior to Jupiter. 

For a moving Jupiter, 
the stress-energy tensor has additional components 
$T^{0i}$, which are proportional to $v^i_J/c$ and generate $v_J/c$ corrections, 
and $T^{ij}$, which are proportional to $v^i_J v^j_J/c^2$ and may be 
neglected because Jupiter's speed is considerably less than the speed of light.
The easiest way to obtain $T^{\mu \nu}$, $R_{\mu \nu}$ and $g_{\mu \nu}$ 
for the non-static case 
is to construct the Lorentz transformation  
that takes a non-moving Jupiter and sends it moving with velocity $\vec v_J$. 
This Lorentz tranformation is then applied to the tensors of the static case. 
Having obtained the metric for a non-static Jupiter, 
one would then need to compute the Shapiro time delay for this case. 

The above procedure for computing the $v_J/c$ corrections 
to the Shapiro time delay difference $\Delta \left( {t_1,t_2} \right)$ 
is, in principle, the one adopted by Kopeikin 
and is quite involved.\ct{kopeikinschafter9902030}  
However, some simplifications occur 
because Jupiter's gravity is weak and $v_J/c$ is small. 
One begins with the Newtonian approximation to Einstein's equations 
and incorporates the $v_J/c$ effects as perturbative corrections, 
which is the commonly used post-Newtonian approximation. 

There is a simpler way to proceed, however, 
and that is to adopt a reference frame in which Jupiter is static. 
In such a frame, 
the Earth moves at a velocity $\vec v_E$ 
given by 
\be
 \vec v_E = - \vec v_J
\,, 
\ee 
while Jupiter's velocity is zero. 
The formula in Eq.\bref{deltalo} 
for the static case is then valid. 
One only needs to incorporate the effects 
of having moving observation stations on Earth 
into the Shapiro time delay difference. 
This procedure is valid because 
(1) Einstein's theory is Lorentz invariant 
and 
(2) during the time 
in which the quasar rays propagate from Jupiter to the Earth, 
Jupiter moves almost in a straight line with constant speed. 
The propagation period is sufficiently short 
that the orbital motion of Jupiter around the Sun is not important. 
Since the same is true for the Earth, 
observers on both planets are essentially inertial 
during the time scales relevant to the experiment. 

Because the Earth is moving, 
the distance $\vec B_{sf}$ between points 1 and 2 
as measured in this static-Jupiter frame 
is not equal to $\vec B$ as measured on Earth. 
In other words, 
if the first quasar signal arrives at $\vec x_1 (t_1)$ at time $t_1$, 
then the Earth will move a short distance 
during the time in which it takes 
the second signal to arrive at $\vec x_2 (t_2)$. 
Place two observers in the static-frame 
(meaning that they are not moving with respect to Jupiter)  
so that one is located at the point $1$ at time $t_1$ 
and the another is at the point $2$ at time $t_2$. 
Then use these observers to make the time measurements. 
Since the situation is completely static, 
the formula for the static case may be used. 

The difference between the times 
at which the two measurements are made is 
\be
 t_2 - t_1 = 
    | \vec x_2 (t_2) - \vec x_0 |/ c - 
    | \vec x_1 (t_1) - \vec x_0 |/ c + 
      \Delta \left( { t_1 , t_2 } \right)
\,. 
\label{deltadefalt}
\ee 
Here, 
$\vec x_0$ is the position of the quasar,  
and 
$ | \vec x_2 (t_2) - \vec x_0 |/ c - | \vec x_1 (t_1) - \vec x_0 |/ c$ 
is the time difference that occurs when gravitational effects 
are absent. 
Eq.\bref{deltadefalt} is an alternative definition 
of $\Delta \left( { t_1 , t_2 } \right)$. 

The leading contribution to this time difference is 
\be
  t_2 - t_1 \approx 
   - { { \vec K \cdot \vec B } \over {c} } 
   + \Delta \left( t_1 , t_2 \right) 
\,, 
\label{t2t1diff}
\ee 
where the first term is, in general, larger than the second 
and arises from the first two terms 
in Eq.\bref{deltadefalt}. 
Here, 
$\vec K$, which is perpendicular to $\vec n$, 
is a unit vector pointing in the direction of the quasar 
as seen from Earth. 
Since during the time $t_2 - t_1$,  
the Earth moves a distance $\vec v_E \left( { t_2  - t_1 } \right)$, 
the displacement between detectors in the static frame $\vec B_{sf}$ 
is not the same as that in the Jupiter-moving frame $\vec B$ 
but the two are related by  
\be
  \vec B_{sf} = \vec B + \vec v_E \left( t_2  - t_1 \right) 
  \approx  \vec B 
   - { { \vec K \cdot \vec B } \over {c} } \vec v_E 
     + \Delta \left( {t_1,t_2} \right) \vec v_E
\,. 
\label{bsfb}
\ee 

The Earth's motion gives rise to three $v_J/c$ effects. 
The first occurs because 
$\vec B_{sf}$ needs to be used in Eq.\bref{deltalo}.
When this substitution is performed, 
a correction of 
$ 
 { { 4G_N M_J \vec n_{sf} \cdot \vec v_E \vec K \cdot \vec B } 
    / ( {\xi c^4} ) }
$ 
is generated.
The effect of the last term 
in Eq.\bref{bsfb} 
may dropped since it is proportional to $G_N^2$.  

The second $v_J/c$ correction occurs 
if the Earth has any motion toward (or away from) Jupiter. 
For example, 
if the Earth were moving toward Jupiter, 
then station 2 would be moving toward the quasar signal 2 
during the time after station 1 had detected signal 1 
but before station 2 had received signal 2. 
Station 2 would then record a smaller time delay 
than if the Earth had not been moving. 
In other words, 
the time delay is reduced (or increased) by an amount
$\delta \Delta \left( {t_1,t_2} \right)$ 
that is equal to the time 
it takes light to travel the distance determined 
by the difference between $\vec B_{sf}$ and $\vec B$. 
The corresponding correction due to the second term  
in Eq.\bref{bsfb}
is independent of Newton's constant and is a contribution to 
the first part of Eq.\bref{deltadefalt} 
that involves detector distance differences. 
The third term in Eq.\bref{bsfb} leads to   
\be 
  \delta \Delta \left( {t_1,t_2} \right) = 
    - { {\vec K \cdot \vec v_E} \over {c} } \Delta \left( {t_1,t_2} \right) 
\,. 
\label{deltadelta}
\ee 

To convert the result to the Jupiter-moving frame, 
one substitutes $\vec v_E = - \vec v_J$. 
Combining the above two effects with the leading term, 
one finds\ct{samuelprl}  
\be
  \Delta \left( {t_1,t_2} \right) = 
    -{ {4 G_N M_J}  \over {\xi c^3} } 
   \left( { 
      \vec n_{sf} \cdot \vec B
      \left( { 1 + { {\vec K \cdot \vec v_J} \over {c} } } \right) + 
      { {\vec K \cdot \vec B \vec n_{sf} \cdot \vec v_J} \over {c} } 
   } \right)
\,. 
\label{deltaintermediate} 
\ee
The above result is written in terms of quantities 
as measured by an observer on Earth with the exception of $\vec n_{sf}$, 
which gives rise to a third effect. 
Because the Earth is moving with respect to the static frame, 
the direction of the quasar as observed in the two frames differ:
$\vec K_{sf} \approx \vec K + \left( { \vec n \cdot \vec v_j / c } \right) \vec n$. 
Since $\vec n$ is defined to be perpendicular to $\vec K$, 
it too differs in the two frames: 
$\vec n_{sf} \approx \vec n - \left( { \vec n \cdot \vec v_j / c } \right) \vec K$. 
When $\vec n_{sf}$ is substituted into Eq.\bref{deltaintermediate}, 
the last term is cancelled:  
\be
  \Delta \left( {t_1,t_2} \right) = 
    -{ {4 G_N M_J}  \over {\xi c^3} }  
      \vec n \cdot \vec B
      \left( { 1 + { {\vec K \cdot \vec v_J} \over {c} } } \right)  
\,. 
\label{deltafinal} 
\ee
As indicated by a lack of subscripts $sf$, 
all quantities are now those measured by an observer on Earth. 
Equation \bref{deltafinal} has no $1/\theta_{obs}^2$ term 
so that there is no enhancement of the velocity-dependent effects.  

The leading term 
for $\Delta \left( {t_1,t_2} \right)$ 
in Eq.\bref{deltafinal} 
is of order $10^{-10}$ seconds, 
well within the measuring capability of the Jupiter/quasar experiment. 
However, 
the $v_J/c$ term is not bigger than $10^{-14}$ seconds, 
which is more than 100 times smaller than 
what was measurable by the VLBI stations. 
Therefore 
reference \ct{fomalontkopeikin} was 
insensitive to the $v_J/c$ term. 
Furthermore, 
this $v_J/c$ correction 
is masked by larger corrections 
that are suppressed by factors such as $B/\xi$ and $\theta_{obs}$ 
compared to the leading term.
The above analysis leaves no doubt 
that Fomalont and Kopeikin did not measure the speed of gravity. 

Here is an example of a correction down by the order of $\theta_{obs}$ 
that can be up to five times bigger than the $v_J/c$ term.
If $\vec B$ has a component in the direction of the quasar, 
then the differences in the distances $r_{1J}$ and $r_{2J}$  
in Eq.\bref{deltalo0} generate the following correction: 
\be 
   {{2 G_N M_J} \over {c^3}}
   { {\vec K \cdot \vec B} \over {R_{EJ} } } 
\,. 
\label{rejcorrection} 
\ee 
It is smaller than the leading term by 
a factor of $0.5 \xi/R_{EJ} \sim 0.5 \theta_{obs}$. 

\medskip
{\bf\large\noindent 4.\ The Kopeikin Formula 
for $ \Delta \left( {t_1,t_2} \right)$}\vglue 0.2cm
\setcounter{section}{4}   
\setcounter{equation}{0}  

Using the post-Newtonian approximation, 
Kopeikin obtained the following result 
for $ \Delta \left( {t_1,t_2} \right)$\ct{kopeikin0212121}
\be
  \Delta \left( {t_1,t_2} \right) = 
  \left( { 1 + { {\vec K \cdot \vec v_J} \over {c}} } \right)
  {{2G_N M_J } \over { c^3}} \ln { 
  \left[ {
      { r_{1J}\left( {s_1} \right) + 
      \vec K\cdot \vec r_{1J}\left( {s_1} \right) }
     \over {r_{2J}\left( {s_2} \right) + 
      \vec K\cdot \vec r_{2J}\left( {s_2} \right)} 
          } \right] }
\,, 
\label{deltakopeikin} 
\ee
where $\vec r_{1J} (s_1) \equiv \vec x_1 - \vec x_J (s_1)$ 
and $\vec r_{2J} (s_2) \equiv \vec x_2 - \vec x_J (s_2)$ 
are respectively the distance vectors 
between the observation points 1 and 2 and Jupiter 
evaluated at the retarded times 
$$ 
 s_1 = t_1 - \vert \vec x_1 - \vec x_J ( {s_1} ) \vert /c
\,, 
$$ 
\be 
 s_2 = t_2 - \vert \vec x_2 - \vec x_J ( {s_2} ) \vert /c 
\,. 
\label{retardedtimes}
\ee 
Note that Kopeikin generally works in a frame in which Jupiter is moving 
and the Earth is not so that Jupiter's position $\vec x_J (t) $ 
varies with time $t$, 
while that of a detector station does not. 

The term 
in Eq.\bref{deltakopeikin} 
involving 
$
   { {\vec K \cdot \vec v_J} / {c}} 
$
is discarded by Fomalont and Kopeikin due to its smallness. 
It is the main $v_J/c$ correction  
in Eq.\bref{deltafinal}. 

At time $s$, 
let $\theta_1 (s)$ (respectively, $\theta_2 (s)$) 
be the angle between $\vec r_{1J} (s)$ (respectively, $\vec r_{1J} (s)$) 
and $\vec K$ (the unit vector pointing toward the quasar). 
Then  
$$
r_{1J}\left( {s} \right) + 
     \vec K \cdot \vec r_{1J}\left( {s} \right) = 
 r_{1J} \theta^2_1 (s) / 2 + O(\theta^4_1 (s)) 
\,,
$$
\be
r_{2J}\left( {s} \right) + 
     \vec K \cdot \vec r_{2J} \left( {s} \right) = 
 r_{2J} \theta^2_2 (s) / 2 + O(\theta^4_2 (s)) 
\,. 
\label{numden} 
\ee 

The times $s_1$ and $s_2$ at which one is to evaluate 
$\vec r_{1J} (s_1)$ and $\vec r_{2J} (s_2)$ 
in Eq.\bref{numden} 
are considerably earlier than the observation times $t_1$ and $t_2$. 
Indeed, 
since 
$| \vec x_1 - \vec x_J \left( {s_1} \right)|/c$ 
is about the time it takes a quasar ray to travel from Jupiter to Earth, 
$s_1$ 
in Eq.\bref{retardedtimes} 
corresponds to when the quasar ray passes near Jupiter. 
The same is true for $s_2$. 
Therefore, 
it is Jupiter's position at this moment that is revelant. 
This is physically reasonable since this is when the planet  
exerts its biggest influence on the rays. 

Combining the results of the previous paragraphs, 
one sees that Eq.\bref{deltakopeikin} 
agrees with Eqs.\bref{deltalo0} and \bref{deltafinal}
since 
$$
  r_{1J} \theta^2_1 (s) \approx \xi_1^2/r_{1J}
\,,  
$$ 
\be
  r_{2J} \theta^2_2 (s) \approx \xi_2^2/r_{2J} 
\,. 
\label{agreement}
\ee  

\medskip
{\bf\large\noindent 5.\ The Faulty Analysis 
of the Jupiter/Quasar Experiment}\vglue 0.2cm
\setcounter{section}{5}   
\setcounter{equation}{0}  

Since Eq.\bref{deltakopeikin} is the leading order result 
in Eq.\bref{deltalo} 
up to unmeasurable corrections, 
and Eq.\bref{deltalo} 
has no $v_J/c$ dependence in it, 
the question arises as to how Fomalont and Kopeikin 
exacted $c_g$ from their data.
Although the details of the analysis have not been revealed, 
I have surmised what transpired 
through their publications   
and through correspondence with Sergei Kopeikin. 

Fomalont and Kopeikin 
express the result for the Shapiro delay time difference 
in Eq.\bref{deltakopeikin} of
\be
    S \left( {s_1 , s_2 } \right) \equiv
 {{2G_N M_J } \over { c^3}} \ln { 
  \left[ {
     { { r_{1J}\left( {s_1} \right) + 
      \vec K\cdot \vec r_{1J}\left( {s_1} \right) }
     \over { r_{2J}\left( {s_2} \right) + 
      \vec K\cdot \vec r_{2J} \left( {s_2} \right) } }
          } \right] }
\,, 
\label{sdelta}
\ee
which is a function of retarded times $s_1$ and $s_2$, 
as a fuction of the observation time $t_1$. 
The leading order contribution to $S \left( {s_1 , s_2 } \right)$ is
\be 
 {{4 G_N M_J } \over { c^3}} 
  \ln { 
  \left[ {
      { \xi_{1}\left( {s_1} \right) }
     \over {\xi_{2}\left( {s_2} \right) } 
          } \right] }
\,. 
\label{slo}
\ee 
If Jupiter is moving toward (or away from) the quasar 
then the distances  
$\xi_{1}\left( {t} \right)$ and $\xi_{2}\left( {t} \right)$ 
between Jupiter and quasar-ray-trajectories
decrease (or increase) with time $t$ 
and are smaller (or larger) if evaluated at the time of observation: 
$$
     \xi_{1}\left( {s_1} \right) \approx 
     \xi_{1}\left( {t_1} \right) + \vec n \cdot \vec v_J { {R_{EJ}} \over {c} } 
\,, 
$$ 
\be
     \xi_{2}\left( {s_2} \right) \approx 
     \xi_{2}\left( {t_1} \right) + \vec n \cdot \vec v_J { {R_{EJ}} \over {c} } 
\,. 
\label{xit1exp} 
\ee 
See Figure 2. 

As is physically clear, 
the difference $\Delta \xi = \xi_{1} (t) - \xi_{2} (t)$ 
does not change with time  
as Eq.\bref{xit1exp} shows. 
However, 
the leading order result 
in Eq.\bref{slo} now does
since the substitution 
in Eq.\bref{xit1exp} leads to 
\be 
   \ln { 
  \left[ {
      { \xi_{1}\left( {s_1} \right) }
     \over {\xi_{2}\left( {s_2} \right) } 
          } \right] } = 
    \ln { 
  \left[ {
      { \xi_{1}\left( {t_1} \right) }
     \over {\xi_{2}\left( {t_1} \right) } 
          } \right] } 
  + { {\vec n \cdot \vec v_J R_{EJ}} \over {c} } 
   \left( { { {1} \over {\xi_1 \left( {t_1} \right)} }  }  -
          { { {1} \over {\xi_2 \left( {t_1} \right)} }  } \right) 
\,. 
\label{lnt1exp}
\ee
The second term 
in Eq.\bref{lnt1exp} is 
\be
     { { \vec n \cdot \vec v_J R_{EJ}} \over {c} } 
   \left( { { {1} \over {\xi_1 \left( {t_1} \right)} }  }  -
          { { {1} \over {\xi_2 \left( {t_1} \right)} }  } \right) \approx 
  - { {\Delta \xi \left( {t_1} \right) } \over {\xi^2 \left( {t_1} \right) } } 
    { {\vec n \cdot \vec v_J R_{EJ}} \over {c} } = 
   { { \vec n \cdot \vec B } \over {\theta_1^2 \left( {t_1} \right) } } 
    { {\vec n \cdot \vec v_J } \over {c R_{EJ}} } 
\,. 
\label{secondterm}
\ee

Summarizing, 
\be 
  S \left( {s_1 , s_2 } \right) - S \left( {t_1 , t_1 } \right) 
  \equiv \Delta_R 
  \approx
  {{4G_N M_J\vec n\cdot \vec B \vec n \cdot \vec v_J } 
    \over { R_{EJ} \theta_1^2 \left( {t_1} \right) c^4}} 
\,. 
\label{deltar}
\ee
The right-hand side of the equation 
contains the artificially $1/\theta^2$ enhancement claimed 
by Kopeikin.\ct{kopeikin0105060,kopeikin0212121} 
It arises because the position of Jupiter changes 
as the quasar signals travel from the Jupiter region to Earth. 
The distance from Jupiter of the (almost) linear path 
of the quasar ray 
when it is detected at a VLBI station 
is different from the distance when it passed by Jupiter. 
Fomalont and Kopeikin used tables to determine Jupiter's position at $t_1$. 
Such information  
allows one to determine $S \left( {t_1 , t_1 } \right)$ analytically. 
So if one experimentally measures 
$S \left( {s_1 , s_2 } \right)$ 
and subtracts $S \left( {t_1 , t_1 } \right)$, 
then one can fit the data to $\Delta_R$  
in Eq.\bref{deltar} 
to extract the artificially generated second term. 

Since $S \left( {s_1 , s_2 } \right)$ 
has no measureable $v_J/c$ dependence in it, 
there is implicit $v_J/c$ dependence in $S \left( {t_1 , t_1 } \right)$. 
This is because 
\be
  S \left( {t_1 , t_1 } \right) \approx 
    -{ {4 G_N M_J \vec n \cdot \vec B }  \over {\theta_1 r_{1J} c^3} } 
\label{s1}
\ee 
can be written as 
\be
  S \left( {t_1 , t_1 } \right) \approx  
    -{ {4 G_N M_J \vec n \cdot \vec B} 
  \over { (\theta_{obs} - { {\vec n \cdot \vec v_J} \over {c}} ) r_{1J} c^3} } 
\label{s1vdep}
\ee
to order $(v_J/c)^2$. 

The Introduction argued that perhaps 
one way to extend Einstein's theory to the $c_g \ne c$ case 
is to replace $1/c$ by $1/c_g$ in the evalution of retarded times. 
Kopeikin does this 
in $S \left( {s_1 , s_2 } \right)$ 
when parametrizing it in terms of $t_1$. 
This, however, is incorrect. 
The difference $\Delta_R$ between 
$S \left( {s_1 , s_2 } \right)$
and 
$S \left( {t_1 , t_1 } \right)$ 
is due to Jupiter's change of position  
during the time it takes quasar rays 
to travel from the vicinity of Jupiter to Earth. 
This depends on the speed of rays, 
which is the speed of light, and not on the speed of gravity. 

Thus, 
when data with error bars are fit 
to $\Delta_R$ 
and the fitting function uses $c_g$ instead of $c$, 
one is guaranteed to obtain the result $c_g \approx c$. 
Fomalont and Kopeikin's announcement that the speed of gravity 
is the speed of light to within 20\% has no content. 

\medskip
{\bf\large\noindent 6.\ Can the Speed of Gravity 
Be Defined for the Jupiter/Quasar Experiment?}\vglue 0.2cm 
\setcounter{section}{6}   
\setcounter{equation}{0}  

This section addresses the theoretical issues raised 
in references \ct{asada,will,faber,asada0308343,carlip}. 
Those references argued 
that Kopeikin's formula for the Shapiro time delay difference 
should involve the speed of light and not the speed of gravity. 
This debate is over how to define the speed of gravity 
in Einstein's general theory of relativity. 

In the static frame, 
Jupiter is not moving. 
The curvature of space-time created by the massive planet 
is governed by Eq.\bref{schwarzschild} and 
is static. 
The effects of gravity are not propagating 
and the speed of gravity concept is non-existent. 
The velocity dependent corrections 
obtained in Eq.\bref{deltafinal} arise 
due to the motion of the VLBI stations 
as they detect the radio waves. 
Since radio waves travel at the speed of light, 
the velocity dependent corrections must be proportional to $v_J/c$, 
and this is evident in the derivation 
of Eq.\bref{deltafinal} 
in Section 3. 
It makes no sense to replace $v_J/c$ by $v_J/c_g$ from the viewpoint 
of the static frame. 

On the other hand, 
in the frame in which Jupiter moves, 
the Introduction suggested that to extend Einstein's theory 
to the case $c_g \ne c$ 
one should replace retarded times as determined by $c$ by retarded times 
as determined by $c_g$. 
It would therefore {\it {seem}} 
as if the speed of gravity concept could be defined 
for this situation. 

However, 
results as measured in one frame must be consistent 
with those measured in another. 
One is forced to conclude that the speed of gravity concept 
is theoretically inconsistent for the Jupiter/quasar experiment. 
If one tries to define the general theory of relativity with $c_g \ne c$ 
for this problem, 
one violates Galilean invariance. 

Originally, 
C.\,Will has suggested that perhaps the speed of gravity might 
enter as a $v_J^2/c_g^2$ effect.\ct{will}  
However, 
the above reasoning still applies. 
In relating the static frame results 
to the Jupiter-moving frame, 
one needs to use Lorentz transformations 
if higher velocity effects are to be considered. 
Therefore, 
an extension of Einstein's theory to describe the Jupiter/quasar experiment
for which the linear velocity corrections 
are of the form $v_J/c$ but the quadratic corrections 
are of order $v_J^2/c_g^2$ for $c_g \ne c$
would violate Lorentz invarince 
(but not Galilean invariance). 
Recently, 
C.\,Will has also come to the same conclusion 
that $c_g$ does not appear 
in any higher power $\left( {v_j/c} \right)^n$ 
correction for a constantly moving Jupiter.\ct{willprivate}  

As mentioned 
in my Physical Review Letter, 
the above argument would fail 
if Jupiter (or another massive object) were accelerating toward (or away from) 
the quasar rays (or other electromagnetic waves). 
It is possible 
that the speed of gravity could be defined for this situation. 
The parameter $c_g$ would then be associated 
with acceleration effects. 
It might be worth analyzing this case as a theoretical possibility. 
Unfortunately, 
it is unlikely that such a system within or beyond the solar system 
exists with sufficiently large effects 
as to be measurable with current VLBI instruments. 

\medskip
{\bf\large\noindent 7.\ The Response by Fomalont and Kopeikin}\vglue 0.2cm
\setcounter{section}{7}   
\setcounter{equation}{0}  

In fairness to Fomalont and Kopeikin, 
it should be said that they have not accepted the conclusions 
of my Physical Review Letter nor the criticisms of others. 
They have continued submitting 
papers\ct{kopeikin0310059,fomalontkopeikin0310065,fomalontkopeikin0311063}  
arguing that they did indeed measure the speed of gravity.
This section addresses those papers. 
Derivations of the formula 
in Eq.\bref{deltakopeikin} 
mostly make up the content of these attempted rebuttals, 
while a few paragraphs are devoted to addressing the criticisms 
of other authors and of my Physical Review Letter. 
These paragraphs contain errors and false statements. 

For example, 
reference \ct{kopeikin0310059} 
says, 
``This part of the experiment was drastically misunderstood by
Samuel who assumed that we measured position of quasar with respect to Jupiter
by measuring the relative position of the quasar with respect to Jupiter in radio.'' 
Similar statements appear 
in reference \ct{fomalontkopeikin0311063} 
(``A fundamental flaw in Samuel's interpretation was his assumption 
that the direction to Jupiter was directly measured by VLBI network 
in the detection experiment so he confused the propagation of gravity 
and the propagation of radio waves.'')
and reference \ct{fomalontkopeikin0310065} 
(``Unfortunately, Samuel incorrectly assumed that the experiment directly compared the
radio position of the quasar with that of Jupiter, and that the direction of Jupiter was
determined by a photon reflected from its surface.'').
My Physical Review Letter never made such statements, 
nor does this review:  
The criticism of the manner in which the data analysis was performed, 
which is presented in Section 5,
focuses on the parametrization of and the expansion about the observation time $t_1$ 
by Kopeikin 
of the Shapiro time delay difference. 

The unpublished work \ct{fomalontkopeikin0310065} continues 
``The experiment monitored the position of the quasar as
a function of the atomic time by the arrival of the quasar's photons at the telescope, 
while the Jupiter's position was determined separately via a precise JPL ephemeris, 
evaluated at the same atomic time as the arrival of a photon 
(via standard transformations from ephemeris time to atomic time). 
Hence the actual angle used for measuring $\Delta$ is
$\theta_1$, not Samuel's $\theta_{obs}$. 
Thus, the $v_J/c$ correction $\Delta_R$ 
was clearly separated from $\Delta_S$ and measured with a precision of 20\%.'' 
Although these statements are intended to underpin my Physical Review Letter, 
anyone who truly understands the situation will realize 
that they actually support the Letter 
and the criticism in Section 5 levelled 
at the data analysis of the Fomalont/Kopeikin measurement. 

Here is example of mistating the work of others 
as a means of defending the theory behind the quasar/Jupiter experiment. 
Reference \ct{fomalontkopeikin0311063} says, 
``Our definition of the "speed of gravity" is more general than 
that used by Asada, Samuel, and Will 
[\ct{asada}, \ct{samuelprl}, \ct{will}] who
limited its meaning as the speed of propagation of gravitational waves.''
and 
``In their formulations of the experiment,
these authors [Asada, Samuel, Will] 
assumed only far-field gravitational effects, 
where gravitational waves are dominant and differentiation
between $c$ and $c_g$ occurs only at orders of $(v/c)^2$ 
beyond Shapiro delay and higher.
This was one reason why the
`speed of light' was interpreted as causing the observed aberration. 
However, the experiment was performed in the
near-field of the quasar radio wave-Jupiter interaction 
where gravitational modes not associated with gravitational waves are dominant.'' 

The work of the three above-cited authors 
focused on the speed of propagation of the effects of gravity 
and {\it {not}} on the speed of gravity waves 
as falsely claimed in these quotes. 
My Physical Review Letter and this review did not discuss 
gravitational waves 
and have emphasized that the Shapiro time delay difference 
is due to relatively {\it {short-distance effects}}; 
I never ``assumed only far-field gravitational effects''.
In actual fact, 
it is Kopeikin's formalism that ends up distorting and mixing up 
the short- and long-distance gravitational effects of Jupiter 
as explained in Section 5. 

Reference \ct{kopeikin0310059} states, 
``The goal of the jovian deflection experiment was to
distinguish two angles $\theta_1$ and $\theta_{obs}$. 
Confirmation that the apparent position of the quasar
in the sky makes the angle $\theta_1$ rather than $\theta_{obs}$ 
with respect to Jupiter is a proof that gravity
propagates with the speed $c_g$.''
This is not true: 
The difference between $\theta_1$ and $\theta_{obs}$ is due to the motion of Jupiter 
during the period in which the quasar waves travel to Earth. 
Quasar waves travel at the speed of light 
so that $c_g$ cannot be extracted from a difference 
of $\theta_1$ and $\theta_{obs}$; 
See Eqs.\bref{thetaobstheta1} and \bref{deltar}
where it is incorrect to replace $c$ by $c_g$.  

Reference \ct{kopeikin0310059} presents another example of faulty reasoning: 
``The equation 
\be
  \Delta \left( {t_1,t_2} \right) = 
    -{ {4 G_N M_J}  \over {c^3 \theta r_{1J} } } 
   \left( { 
      \left( { 1 + { {\vec K \cdot \vec v_J} \over {c_g} } } \right) \vec n \cdot \vec B + 
      \vec K \cdot \vec B \left( {
           { { \vec n \cdot \vec v_E} \over {c} } 
         - { { \vec n \cdot \vec v_J} \over {c_g} } 
                         } \right)
   } \right)
\label{7p5} 
\ee 
does not contain terms being quadratic in $1/\theta$. 
It may make an impression that the orbital motion of Jupiter 
does not provide any significant deviation from the Einstein's prediction 
of the light deflection because all velocity-dependent
terms in the right side of Eq.\bref{7p5} are smaller than the main term 
(proportional to $(\vec n \cdot \vec B)/\theta$) 
by a factor of $10^{-4}$ and can not be observed with the present-day technology.
This was the reason for Samuel's statement that terms of order $v/c$ beyond the
Shapiro time delay are not observable. 
This statement is erroneous because the Shapiro time delay must be calculated 
in terms of the present position of Jupiter at the time
of observation $t_1$.''

The last statement is incorrect. 
Kopeikin's own formalism (the right-hand side of Eq.\bref{deltakopeikin}) 
leads to the conclusion  
that the Shapiro time delay difference $\Delta$ should 
be computed at the retarded time, 
which is near the time at which the quasar rays pass Jupiter. 
Furthermore, 
it is not necessary to evaluate $\Delta$ at $t_1$:  
Since the Shapiro time delay difference is generated 
when the rays are in the vicinity of Jupiter, 
$t_1$, 
which determines the retarded time $s_1$ 
in Eq.\bref{retardedtimes}, 
may be evaluated at {\it {any}} time after the rays 
have passed well beyond Jupiter. 
As is physically clear, 
tens of millions of kilometers beyond Jupiter, 
the Shapiro time delay difference will have been almost 100\% generated 
and then remain essentially unchanged. 
As is evident from Eqs.\bref{deltalo0} and \bref{rejcorrection}, 
there is only very weak dependence of $\Delta$ 
on the Earth-Jupiter distance $R_{EJ}$.  

In both references \ct{kopeikin0310059} and \ct{fomalontkopeikin0311063}, 
Kopeikin claims that his formalism is Lorentz invariant, 
which is not the case. 
In Section 6, 
we argued that if $c_g \ne c$ 
then not only is Lorentz invariance violated 
for the Jupiter/quasar experiment 
but also Galilean invariance. 

Kopeikin derived Eq.\bref{7p5}  
in part
to show that his formalism agreed with 
the result in my Physical Review Letter 
when $c_g = c$. 
Eq.\bref{7p5} is Kopeikin's generalization 
to arbitrary values of the speed of gravity. 
In this equation, 
$\vec v_E$ and $\vec v_J$ 
are the velocities of the Earth and Jupiter 
as determined in a coordinate system in which the Sun is at rest. 
However, 
Eq.\bref{7p5} does not depend on $\vec v_J - \vec v_E$ 
as required by Galilean invariance. 
The frame-dependence is manifest: 
Hypothetically place the Sun at another location, 
give it a velocity of $\vec v_S$ relative to the original static Sun, 
have the Earth and Jupiter move 
with the same velocities that they originally had in the Jupiter/quasar experiment, 
then Eq.\bref{7p5} becomes 
$$
  \Delta \left( {t_1,t_2} \right) = 
$$
\be
    -{ {4 G_N M_J}  \over {c^3 \theta r_{1J} } } 
   \left( { 
      \left( { 1 + { {\vec K \cdot (\vec v_J - \vec v_S) } \over {c_g} } } \right) 
                \vec n \cdot \vec B + 
      \vec K \cdot \vec B \left( {
          { { \vec n \cdot (\vec v_E - \vec v_S) } \over {c} }
        - { { \vec n \cdot (\vec v_J - \vec v_S) } \over {c_g} } 
                         } \right)
   } \right)
\label{sunmotion}
\ee 
because, with respect to the ``new'' Sun, 
Jupiter and the Earth now move with velocities 
$\vec v_J - \vec v_S$ and $\vec v_E - \vec v_S$. 
The dependence on the ``new'' Sun's velocity $\vec v_S$ does not drop out. 
How can the Shapiro time delay difference due to Jupiter 
and observed on Earth be so dependent on the Sun's motion?
The answer is that Eq.\bref{7p5} is wrong. 
In a frame in which both the Earth and Jupiter move, 
the correct result is Eq.\bref{deltafinal} with $\vec v_J$ 
replaced by $\vec v_J - \vec v_E$. 

Reference \ct{fomalontkopeikin0311063} 
emphasizes incorrectly that it is the postion of Jupiter 
at the time of observation that is relevant: 
``the gravitational force, 
acting on photons as a space-like vector, 
is sensitive to the present position of Jupiter.''
For the sake of argument, 
let us assume that this is the case. 
When Kopeikin expands 
about the time of observation $t_1$, 
he obtains, 
for the case $c_g = c$,  
a $v_J/c$ correction of\ct{kopeikin0105060} 
\be
  \Delta \left( {t_1, t_2} \right) = 
    -{ {4 G_N M_J}  \over {\theta_1^2 r_{1J} c^4} } 
   \left( { \vec B \cdot \vec v_J - \vec K \cdot \vec v_J \vec K \cdot \vec B } \right)
\,. 
\label{deltarwrong} 
\ee 
It is easy to see that this equation leads to physically unreasonable results. 
Suppose, for example, that Jupiter is moving toward the quasar rays. 
Hypothetically place a second planet twice as far away as Earth is to Jupiter 
and put VLBI stations on it 
along the same lines of observation as determined by quasar ray trajectories 
that pass through VLBI stations on Earth. 
Then one would expect to measure to a high precision 
the same Shapiro time delay difference. 
However, Eq.\bref{deltarwrong} predicts that more than twice 
the Earth-based correction will be measured. 
This follows because 
$1/(\theta_1^2 r_{1J}) = r_{1J}/\xi_1^2 (t_1)$ where $\xi_1 (t_1)$ is 
the distance between Jupiter and the worldline of the quasar ray $1$
at the time of observation $t_1$. 
In making the observations ``downstream'' on the hypothetical planet, 
$r_{1J}$ is doubled and $\xi_1 (t_1)$ is smaller if Jupiter is moving toward the rays 
because the measurement is made later compared to the Earth-based one. 
How can Jupiter, which is so far away, 
have such a dramatic long-ranged influence on $\Delta$? 

The answer is that $v_J/c$ corrections are not given 
by Eq.\bref{deltarwrong}; The correct result is given 
in Eq.\bref{deltafinal}. 
Kopeikin should not have performed 
an expansion about the observation time that separated 
the leading Shapiro time difference result into two pieces. 
As explained in Section 5, 
his leading order result actually has implicit $v_J/c$ dependence in it
(because it is written in terms of $\theta_1$ instead of $\theta_{obs}$ or $\xi$), 
which, when combined with his purported $v_J/c$ correction, 
gives the $v_J/c$-independent result 
of Eq.\bref{deltalo} to order $v_J^2/c^2$. 
This leading order term 
does not depend on how far ``downstream'' the measurement is made. 
The same is true of the $v_J/c$ correction 
in Eq.\bref{deltafinal}. 

\medskip
{\bf\large\noindent 8.\ Summary}\vglue 0.2cm
\setcounter{section}{8}   
\setcounter{equation}{0}  

This work provides the leading $v_J/c$ corrections 
to the Shapiro time delay difference 
and shows that they do not correspond to the ones 
used in the data analysis by Fomalont and Kopeikin 
even when the speed of gravity parameter $c_g$ 
is set equal to $c$. 
The error made by Kopeikin is that he
separated the leading $v_J/c$-independent term 
into two pieces by expanding about the time of observation $t_1$
through the subtraction procedure involving $\Delta_R$ 
in Eq.\bref{deltar}. 
This introduced an artificially $v_J/c$-dependent term. 
It is enhanced by $1/\theta_1$ and 
is an artefact of the $t_1$ expansion. 
Although this $v_J/c$-dependent term should depend 
the speed of light $c$ 
because it is related to the change in position of Jupiter 
during the time in which radio waves travel to Earth, 
Kopeikin incorrectly replaces $c$ by $c_g$. 
Hence, 
when data is parametrized in this way, 
one is guaranteed to obtain a result for $c_g$ that is the speed of light 
to within experimental errors. 
Finally, 
when $c_g \ne c$ 
we agree with others\ct{asada,faber,asada0308343,carlip} 
that the theoretical formalism\ct{kopeikinschafter9902030,fomalontkopeikin0206022,kopeikin0212121,kopeikin0302462,kopeikin0310059,fomalontkopeikin0310065,fomalontkopeikin0311063} 
of Kopeikin and his co-workers  
is flawed when applied to Jupiter/quasar system 
and demonstrated this by showing 
that not only does it violate Lorentz invariance 
but also Galilean invariance. 
While the Jupiter/quasar measurement 
was an extraordinary experimental undertaking 
and a fine achievement in precision, 
it had nothing to do with the determination of the speed of gravity.

\medskip

{\bf\large\noindent Acknowledgments}

I thank Professor Clifford Will 
for some useful comments and suggestions. 
This work is supported in part
by the Director, Office of Science, 
Office of High Energy and Nuclear Physics, 
of the Department of Energy 
under contract number DE-AC03-76SF00098.
Correspondence with Sergei Kopeikin concerning many 
issues is acknowledged.

\bigskip

\def\PRL#1#2#3{ {\it {Phys.{\,}Rev.{\,}Lett.{\,}}}{\bf {#1}}, {#2} ({#3})}
\def\AJ#1#2#3{ {\it {Astrophys.{\,}J.{\,}}}{\bf {#1}}, {#2} ({#3})}
\def\AJL#1#2#3{ {\it {Astrophys.{\,}J.{\,}Lett.{\,}}}{\bf {#1}}, {#2} ({#3})}

\medskip 
{\bf\large\noindent Figure Captions}  

\noindent
Figure 1. The Expected Effect of the Speed of Gravity on the Gravitational Force 
Due to the Motion of Jupiter. \\ 
Jupiter, shown as an open circle, 
is assumed to be moving toward the electromagnetic wave. 
In the top figure, 
$c_g = \infty$ 
and it is the instantaneous position of Jupiter at time $t$,   
which is given by the shaded circle, that is relevant. 
In the middle figure, 
$c_g = c$ 
and the relevant distance is farther away:  
Since it takes time for the gravitational influence of Jupiter to propagate 
to the electromagnetic wave, 
it is Jupiter's position at an earlier time  
that is relevant. 
In the lower figure, 
$c_g < c$.

\medskip 

\noindent
Figure 2. The Definitions of Various Quantities Relevant 
in the Jupiter/Quasar Experiment. \\
The diagram is not drawn to scale for reasons of clarity,  
and, in particular, angles are much larger than in the actual experiment.

\vfill\eject
\end{document}